\documentclass[pss]{wiley2sp}
\pdfoutput=1
\usepackage{amsmath,cite}
\usepackage[colorlinks,plainpages=false,linkcolor=blue,urlcolor=blue,citecolor=blue,pdfpagemode=UseNone,pdfstartview=FitBH]{hyperref}

\begin{document}
\title{Impurity effects on spin dynamics in magnetic and superconducting iron pnictides and chalcogenides}
\titlerunning{Impurity effects on spin dynamics in iron pnictides and chalcogenides}

\author{%
 M.\,A.\,\,Surmach\textsuperscript{\textsf{\bfseries 1}},
 P.\,\,Y.\,\,Portnichenko\textsuperscript{\textsf{\bfseries 1}},
 J.\,T.\,\,Park\textsuperscript{\textsf{\bfseries 2}},
 J.\,A.\,\,Rodriguez-Rivera\textsuperscript{\textsf{\bfseries 3,4}},
 D.\,L.\,Sun\textsuperscript{\textsf{\bfseries 5}},
 Y.~Liu\textsuperscript{\textsf{\bfseries 5}},
 C.\,T.~Lin\textsuperscript{\textsf{\bfseries 5}},
 D.\,S.\,\,Inosov\textsuperscript{\Ast,\textsf{\bfseries 1}}
 }

\authorrunning{M.~A.~Surmach \textit{et al.}}

\mail{e-mail
 \textsf{dmytro.inosov@tu-dresden.de}, Phone: +49-351-463-34731, Fax: +49-351-463-37173\vspace{-5pt}}

\institute{%
 \textsuperscript{1}\,Institut f{\"u}r Festk{\"o}rperphysik, TU Dresden, D-01069 Dresden, Germany\\
 \textsuperscript{2}\,Heinz Maier-Leibnitz Zentrum (MLZ), TU M{\"u}nchen, D-85747 Garching, Germany\\
 \textsuperscript{3}\,Materials Science and Engineering, University of Maryland, College Park, Maryland 20742, USA\\
 \textsuperscript{4}\,NIST Center for Neutron Research, National Institute of Standards and Technology, Gaithersburg, Maryland 20899, USA\\
 \textsuperscript{5}\,Max-Planck-Institut f\"ur Festk\"orperforschung, Heisenbergstra{\ss}e 1, 70569 Stuttgart, Germany}

\received{25 March 2016, revised 29 August 2016, accepted 18 November 2016\smallskip\\Feature article for the Special Issue on Iron-Based High Temperature Superconductors}
\published{as \href{http://dx.doi.org/10.1002/pssb.201600162}{doi:10.1002/pssb.201600162} (Phys. Stat. Sol. B, in press)}


\abstract{%
\abstcol{
In this paper we summarize the effects of magnetic and nonmagnetic impurities on the spin dynamics in Fe-based superconductors and their parent compounds. The effects of chemical substitution, vacancies, and disorder on the suppression or stabilization of superconductivity and spin-density-wave phases are reviewed in the context}{of recent neutron-spectroscopy measurements of spin excitations. We also present new results on the structure of magnetic fluctuations in BaFe$_2$As$_2$ single crystals doped with Mn local moments and discuss them in relationship to the previously reported $(\pi,\pi)$ branch of checkerboard magnetic excitations.
}}

\maketitle

\section{Introduction}

Superconductivity (SC) can be induced in a wide group of materials by chemical substitution or introduction of structural defects. The resulting changes in the ground state are typically caused by a combination of charge-doping, chemical-pressure, and impurity-scattering effects, which are often difficult to separate. While charge-doping effects can be minimized by isovalent substitution or introduction of structural defects, even in these cases more subtle changes of the Fermi surface cannot be avoided completely, as the end member compounds of the substitution series do not have identical electronic structures. These changes become especially important in the proximity to Lifshitz transitions, where the density of states can no longer be considered constant near the Fermi level. Therefore, separating the effect of impurities from influences of chemical pressure and changes in the band structure represents a challenge that can only be solved by a systematic comparison of multiple experiments with predictions of theoretical models.

In early days, impurity effects were investigated in classical experiments on Bardeen-Cooper-Schrieffer (BCS) superconductors \cite{Michael, Muller_Hartmann} to study the nature of the pairing state. It was shown that magnetic impurities are not time-reversal invariant and hence have a drastic effect on superconductivity by breaking Cooper pairs \cite{Anderson_1959, Abrikosov_Gorkov_1959_1, Abrikosov_Gorkov_1959_2, Abrikosov_Gorkov_1959_3}. The following work \cite{Abrikosov_Gorkov_1961} predicted a decrease of the critical temperature ($T_\text{c}$) and energy gap with impurity concentration. On the other hand, it was shown that nonmagnetic impurities (NMI) should not affect directly the physical properties of a conventional superconductor \cite{Anderson_1959}. Further theories predicted the presence of impurity bands within the superconducting energy gap \cite{Shiba_1968, Rusinov_1969}, which were later demonstrated in tunneling experiments \cite{Tsang_1980, Bauriedl_1981}.

In the following decades, the role of impurities for the superconducting phase was studied in non-BCS superconductors, which display contrasting properties to the conventional (BCS) superconductors. Well-studied examples of such unconventional superconductors are high-$T_{\rm c}$ cuprates and heavy-fermion systems \cite{Hotta, Ishida, Steglich}, where SC emerges from a competing magnetically ordered phase. The presence of sign-changing SC pairing in such superconductors makes them much more sensitive to nonmagnetic defects \cite{Mackenzie_1998, Xiao_1988}. In copper oxides SC can be induced only by electron and hole doping into the nearly perfect CuO$_2$ plane, whereas any chemical substitution at the Cu site by other elements dramatically suppresses SC \cite{Kastner_1998, Kivelson_2003, Armitage_2010}. This picture is completely different in the family of Fe-based superconductors (FBS), where Fe-site doping can either enhance or suppress the superconducting state depending on the specific material and the dopant.

The effect of both magnetic and nonmagnetic impurities in parent compounds of iron pnictides has been intensively studied theoretically \cite{LiGuo16, GastiasoroBernardini16}. It was shown that the introduction of a nonmagnetic atom instead of Fe can change the magnetic ground state of the material and stabilize static magnetism \cite{GastiasoroHirschfeld13} or lead to the formation of an anticollinear magnetic order \cite{Weber_Mila_2012}. It was also demonstrated that magnetic impurities can exhibit cooperative behavior due to the Ruderman-Kittel-Kasuya-Yosida interaction mediated by conduction electrons \cite{GastiasoroAndersen13}, which leads in the case of Mn-substituted BaFe$_2$As$_2$ to a new type of short-range checkerboard ($\pi$, $\pi$) fluctuations concomitant with the conventional ($\pi$, 0) stripe-like magnetic order \cite{Tucker_Pratt_2012}.

In superconducting compounds, the investigation of impurity effects provides indirect information about the symmetry of Cooper pairs by studying the suppression of SC as a function of impurity concentration \cite{GolubovMazin97, SchillingBaumgartner16}. According to Anderson's theorem \cite{Anderson_1959, Abrikosov_Gorkov_1959_1, Abrikosov_Gorkov_1959_2, Abrikosov_Gorkov_1959_3}, NMIs act as strong scattering centers and lead to a suppression of SC by pair breaking in the case of an anisotropic gap, for instance for $s_\pm$-wave and $d$-wave gap symmetries. However, in the underdoped regime of FBS, the $s_\pm$ superconducting state can be also enhanced, contrary to this conventional wisdom, due to the suppression of the competing itinerant magnetic order \cite{FernandesVavilov12}. On the other hand, the isotropic $s$-wave superconductors are expected to be sensitive to magnetic and resistant against nonmagnetic impurities. Such distinction in the impact of impurities or controlled disorder, when observed experimentally on a new type of superconductor, gives a tentative hint on the underlying pairing symmetry and superconducting mechanism \cite{WangKreisel13, LiGuo16}.

\begin{figure}[ht]
\includegraphics[width=\columnwidth]{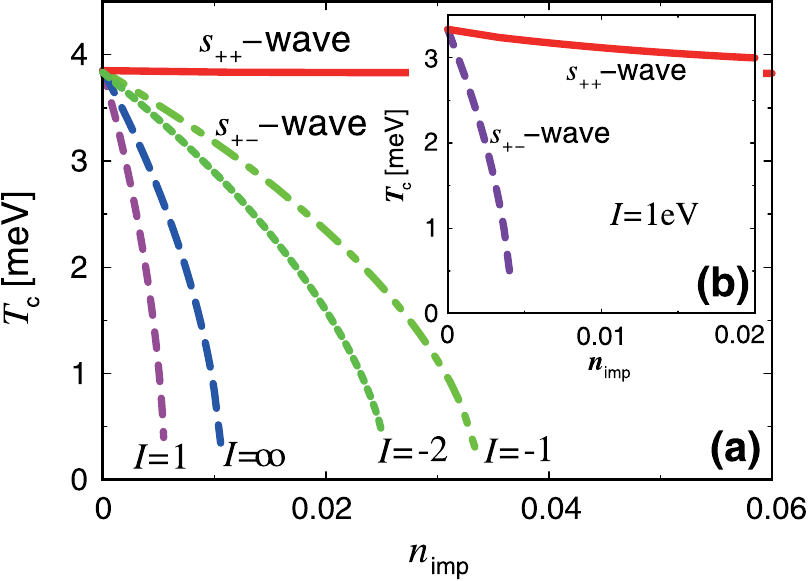}
\caption{Dependence of $T_\text{c}$ on the local impurity concentration ($n_\text{imp}$) in iron pnictides for different pairing symmetries. Colored curves show models for different impurity potentials ($I$). The inset shows models for the pairing symmetry with the largest value of $I$. For undoped model $T_{\text{c0}}=46$\,K. Reproduced from Ref.~\cite{Onari_Kontani_2009}, copyright by the American Physical Society. }
\label{fig:Onari}
\vspace{-0.2em}
\end{figure}

The relation between $T_\text{c}$ and the impurity level for FBS is shown in Fig.\,\ref{fig:Onari} \cite{Onari_Kontani_2009}. These results were based on the five-orbital model given in Ref.~\cite{Kuroki}. The plot shows results for both $s_{++}$-wave and $s_\pm$-wave symmetries for different impurity potentials. It also illustrates the fragility against impurities of the $s_\pm$-wave and the stability for $s_{++}$-wave superconducting states. Authors suggested that in the sign-reversing $s$-wave state, the interband impurity scattering is promoted by the $d$-orbital degree of freedom. At the same time only a comprehensive analysis of different techniques in addition to the $T_\text{c}$ suppression can mark out the exact host compound.

The class of FBS includes different groups of compounds, which significantly differ by their crystal structure, chemical composition, magnetic structure, and superconducting properties. Despite such a variety of features, including pairing symmetry and gap structure, the source of the pairing interaction in all groups is now generally considered to be the same \cite{Hirschfeld_2011}. From the theoretical point of view, two different scenarios, involving spin and orbital fluctuations, were suggested to describe SC in FBS. Complexities in the isolation of impurity effects still raise a discussion of symmetry and structure of the order parameters in some FBS \cite{Hirschfeld_2011}. Due to the multiband structure of the Fermi surface, sign-changing order parameters with an $s_\pm$-wave and $d$-wave symmetry were theoretically proposed \cite{Mazin_2008, Yao_2009, Korshunov_2014}. Such sign-reversing pairing was indirectly confirmed by the presence of the resonance peak in inelastic neuron scattering (INS) experiments \cite{Christianson_2008, Dai_2009, InosovReview16} and later by scanning tunneling spectroscopy (STS) measurements \cite{Yang_Wang_2013}. It was also demonstrated that coexistence of a long-range magnetic order and SC, that is distinctive for several families, is possible only for $s_\pm$ pairing \cite{Vorontsov_2010, Fernandes_2010}. The sign-reversal $s_\pm$-wave symmetry was supported in Fe(Se,Te) compounds by studying quasi-particle interference patterns in the superconducting state by STS \cite{Hanaguri_2010}. A more direct proof of a sign reversing pairing symmetry is expected from future phase-sensitive tests based on combinations of tunnel junctions and point contacts but so far such experiments have only been proposed theoretically \cite{Golubov_2013}. Moreover, other possible candidates for the pairing symmetry have been suggested in FBS from $s_{++}$-wave \cite{Tsai_Zhang_2009, Yao_Chen_2012} to $d$-wave \cite{Yang_Wang_2013, Zhou_Hu_2009} as well as the novel orbital-antiphase symmetry \cite{Yin_Haule_2014}. It is also expected that introduction of disorder may trigger a transition between different pairing symmetries \cite{EfremovKorshunov11, EfremovGolubov13}.

With only a few exceptions like LiFeAs \cite{Pitcher_2008, Tapp_2008}, KFe$_2$As$_2$ \cite{Lee_Kihou_2011}, and Ca$_{10}$(Pt$_4$As$_8$)(Fe$_2$As$_2$)$_5$ (also called 10-4-8) \cite{Ni_Allred2011, SturzerSturzer16}, superconductivity in FBS emerges upon doping from an antiferromagnetic metallic parent compound and may coexist with static magnetic order. Superconductivity, for most FBS, can be tuned by charge doping that suppresses antiferromagnetic (AFM) order. Moreover, a theoretical work \cite{Lv_Moreo_2014} suggested that the coexistence of static or fluctuating magnetic phases with SC has a strong effect on the neutron resonant mode, a ubiquitous feature that is present in most unconventional superconductors \cite{Christianson_2008, Dai_2009, InosovReview16}. Therefore it is important to determine the microscopic origin and evolution of magnetic fluctuations in parent compounds of FBS to better understand the mechanisms of high-temperature SC.

To avoid confusion, in this paper we will describe wave vectors and all reciprocal-space coordinates ($H K L$) in the unfolded notation corresponding to the Fe sublattice (one Fe atom per unit cell) and express them in reciprocal lattice units $1 {\rm r.l.u.} = 2\pi/(a/\sqrt{2})$, where $a$ is the lattice constant of the FeAs plane, so that $a/\sqrt{2}$ is the nearest-neighbor Fe-Fe distance.

Although undoped stoichiometric parent compounds of FBS may exhibit considerably different static AFM arrangements and N\'{e}el temperatures \cite{Dai_2009}, $T_\text{N}$, magnetic interactions are generally considered to be important for superconductivity in nearly all of these materials. The in-plane collinear long-range AFM structure for different iron-pnictide superconductors is identical for all materials, but the out-of-plane component is material dependent. Most iron pnictides share the collinear C-type AFM structure \cite{Wollan_1955}, but some related compounds have G- \cite{Singh_2009} or even an A-type \cite{Quirinale_2013} structure. Another distinct type of collinear structure was found in LaFeAsO$_{1-x}$H$_x$ \cite{Hiraishi_2014}.

At the same time the magnetic structure of Fe-chalcogenides Fe$_{1+x}$Te with $x\!\leq\!9$\% differs from that of Fe-pnictides and represents a commensurate bicollinear spin structure that is rotated by $45^{\circ}$ \cite{Li_2009} with respect to the AFM ordering vector of most Fe-pnictides (1/2 0 1/2), and exhibits incommensurate helical magnetic order which competes with commensurate antiferromagnetism for $x\geq12$\% \cite{Rodriguez_2010, Rodriguez_2011, Rodriguez_2013}. The end member of the Fe$_{1+\delta}$Te$_{1-x}$Se$_x$ series is FeSe, which has no long-range magnetic order \cite{Grechnev_Panfilov_2012}.

The feature of alkali-metal selenides $A_x$Fe$_{2-y}$Se$_2$ ($A$ = K, Rb, Cs, Tl) is the presence of a crystallographic $\sqrt{5}\times\sqrt{5}$ superstructure of Fe vacancies \cite{Zavalij_Bao_2011,Pomjakushin_Sheptyakov_2011} and the coexistence of two spatially separated phases: (i) the majority nonsuperconducting vacancy-ordered AFM phase with magnetic moments aligned along the $\mathbf{c}$ axis; and (ii) the minority paramagnetic phase, which exhibits a superconducting response \cite{Li_Ding_2012,Shermadini_Luetkens_2012, Charnukha_Cvitkovic_2012}. The detailed structure of the superconducting phase was studied by a variety of experimental methods such as: x-ray and neutron diffraction \cite{Ricci_Poccia_2011, Bosak_Svitlyk_2012, Carr_2014}, photoemission \cite{Zhang_Yang_2011, Chen_Xu_2011}, NMR \cite{Kotegawa_Tomita_2012, Texier_Deisenhofer_2012}, and neutron spectroscopy \cite{Friemel_Liu_2012, Friemel_Park_2012}. This phase has reportedly no vacancies in the FeSe layer and an alkali-metal deficient composition $A_x$Fe$_2$Se$_2$, where the actual amount of dopant depends on the sample and experimental method and varies in the range of $0.3\leq x\leq0.6$ \cite{Carr_2014,Texier_Deisenhofer_2012}.

\begin{figure}[b!]
\includegraphics[width=\columnwidth] {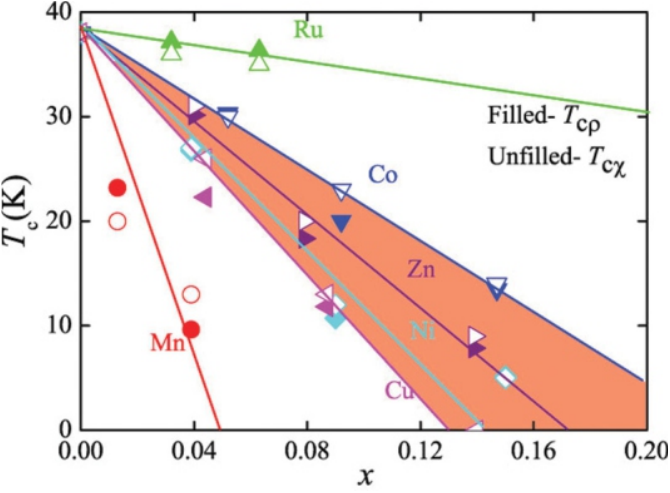}
\caption{The dependence of $T_\text{c}$ suppression in Ba$_{0.5}$K$_{0.5}$Fe$_{2-2x}M_{2x}$As$_2$ by substitution of magnetic and nonmagnetic transition metals ($M$ = Mn, Ru, Co, Ni,
Cu, and Zn). The solid markers represent $T_\text{c}$ obtained from resistivity measurements, whereas open markers show $T_\text{c}$ from magnetization. The superconductivity is suppressed much slower by Ru impurities than by Co, Zn, Ni, Cu, and Mn. Among these impurities Mn is observed as having the strongest suppression effect. The negligible suppression effect from Ru in the present compound is consistent with the `1111' system \cite{Sato_Kobayashi_2010}. The stronger $T_\text{c}$ suppression effects from Mn, Co, Ni, Cu, and Zn are considered to be a result of localization rather than the pair-breaking effect in the s$_\pm$-wave model. Copyright by the American Physical Society. Reprinted from \cite{Li_Guo_2012}.}
\label{fig:Li}
\vspace{-0.2em}
\end{figure}

In iron-pnictide superconductors the static magnetic moment and $T_\text{N}$ gradually decrease with electron doping \cite{Lester_2009}. Contrary to electron doping, hole doping does not separate the structural and magnetic phase transitions, the coexistence of superconductivity with the AFM phase is also present.

As follows from reported experimental and theoretical studies, the introduction of impurities or disorder gives us an instrument for studying of new magnetic states in FBS phase diagrams as well as the competition between magnetism and SC. Our paper gives a short review of the investigations of impurity effects in FBS and their parent compounds. Such effects were intensively studied since conventional superconductors, but there has been no comprehensive review which would elucidate the influence of impurities on magnetic fluctuations in magnetic and superconducting iron arsenides and chalcogenides. The main purpose of this article is to describe the most recent results, to summarize the current understanding of how impurities affect the magnetic structure and spin excitations, as well as the rise and suppression of SC in FBS. Although we mainly focus on the results of neutron spectroscopy measurements, we will also cover important results of other scattering techniques.

\section{The effect of nonmagnetic impurities}

Single NMI has a filled $d$ shell and hence works as a scattering center, resulting in quasiparticle interference patterns (QPI). Any sort of impurity in a metal can be screened by the conduction electrons which leads to Friedel oscillations of the charge and spin density around the defect. A useful technique to map out the variations in electron density near the Fermi level resulting from such oscillations is STS.

Recently, QPI on Cu impurities was studied in Na(Fe$_{0.97-x}$Co$_{0.03}$Cu$_x$)As by STS \cite{Yang_Wang_2013}. Authors considered that Cu plays the role of a weak magnetic impurity compared with the strong magnetic host ions Fe$^{2+}$. The local density of states around the Cu impurity exhibits a systematic evolution, which is well consistent with the coherence length. Therefore Cu impurity is considered as a source of disorder leading to a decay of the in-gap quasiparticle states and consequently inducing Cooper-pair breaking in the strongly anisotropic s$_\pm$ pairing state.

\begin{figure}[t!]
\includegraphics[width=\columnwidth]{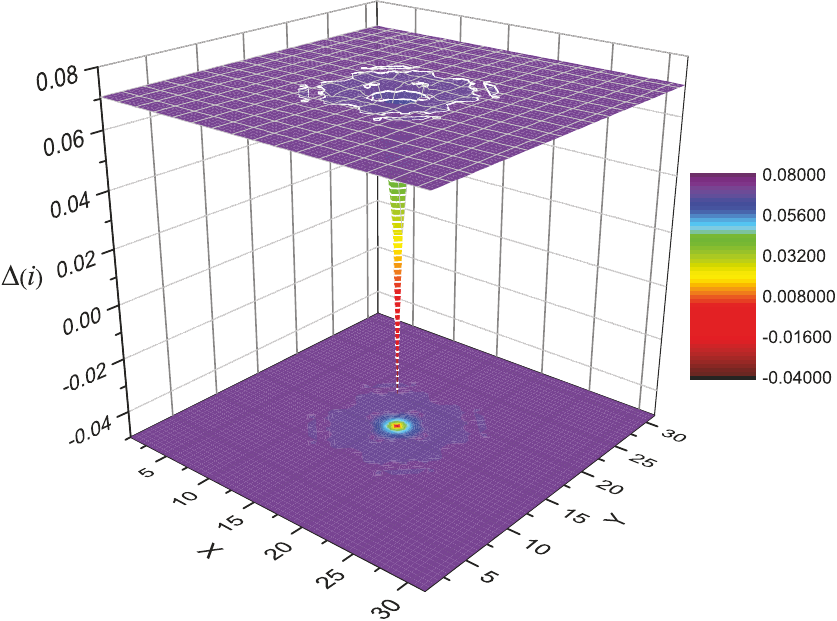}
\caption{The spatial dependence of the SC order parameters $\Delta(i)$ in the presence of a single nonmagnetic impurity reprinted from Ref.~\cite{Yu_Liu_2015}, copyright by Elsevier publishing company.\vspace{-3pt}}
\label{fig:NMI}
\vspace{-0.2em}
\end{figure}

Latest results in Ref.~\cite{Li_Ji_2015} demonstrated the local destruction of SC by the presence of nonmagnetic Zn impurities in the single-crystal nanobridges of  Ba$_{0.5}$K$_{0.5}$Fe$_2$As$_2$ (BKFA), which were studied in the approach of a quasi-one-dimensional superconducting system. Authors observed phase-slip phenomena, revealed in transport measurements, and studied both in-plane and out-of-plane effects of Zn ions on superconductivity. In cuprates, Zn impurities can exclude on a local scale the supercurrent within an area of $\pi\xi^2_{ab}$, where $\xi_{ab}$ is the in-plane coherence length of the superconductor \cite{Pan_Hudson_2000, Nachumi_Keren_1996, Adachi_Yairi_2004}, and result in a two-dimensional (2D) ``Swiss cheese''-like model of the supercurrent distribution. The same model was extended to the third dimension along the $c$ axis (stacks of 2D model separated by Ba/K barrier layer) in 6\% Zn doped BFKA \cite{Li_Ji_2015}, which shows in contrast to previous results \cite{Li_Yuan_2013} that a small amount of Zn impurities leads to the inhibition of SC. Authors estimated the in-plane $\xi_{ab}=2.05$\,nm and out-of-plane $\xi_{c}=1.2$\,nm coherence lengths. The presence of impurities indicates the increase of anisotropy in comparison with previously reported $\xi_{ab}=1.2$\,nm and $\xi_{c}=0.45$\,nm results \cite{Welp_Mu_2009} for the similar compound Ba$_{0.6}$K$_{0.4}$Fe$_2$As$_2$ in the absence of Zn doping.

\begin{figure}[b!]
\includegraphics[width=\columnwidth]{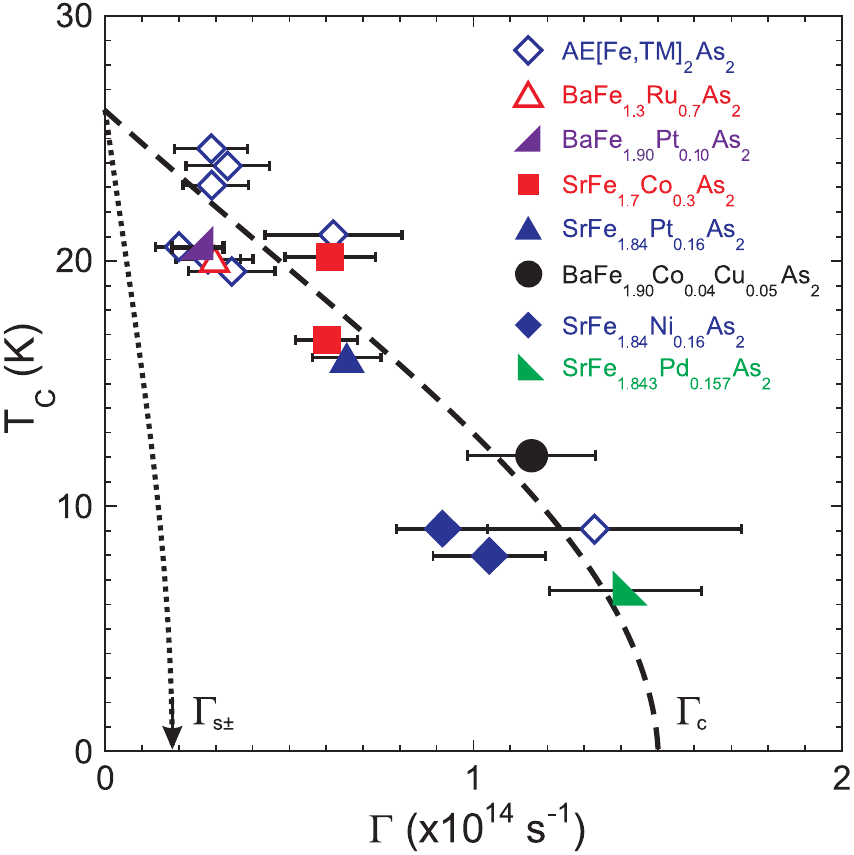}
\caption{Universal correlation between the $T_{\rm c}$ suppression and the transport scattering rate, evidencing the dominant role of impurity scattering for the pair-breaking mechanism in `122'-type superconductors. This effect is much weaker than expected for an $s_\pm$ multiband superconductor with interband scattering (dotted line). Copyright by the American Physical Society. Reproduced from Ref.~\cite{KirshenbaumSaha12}.\vspace{-5pt}}
\label{fig:Kirshenbaum}
\end{figure}

\begin{figure}[t!]
\includegraphics[width=\columnwidth]{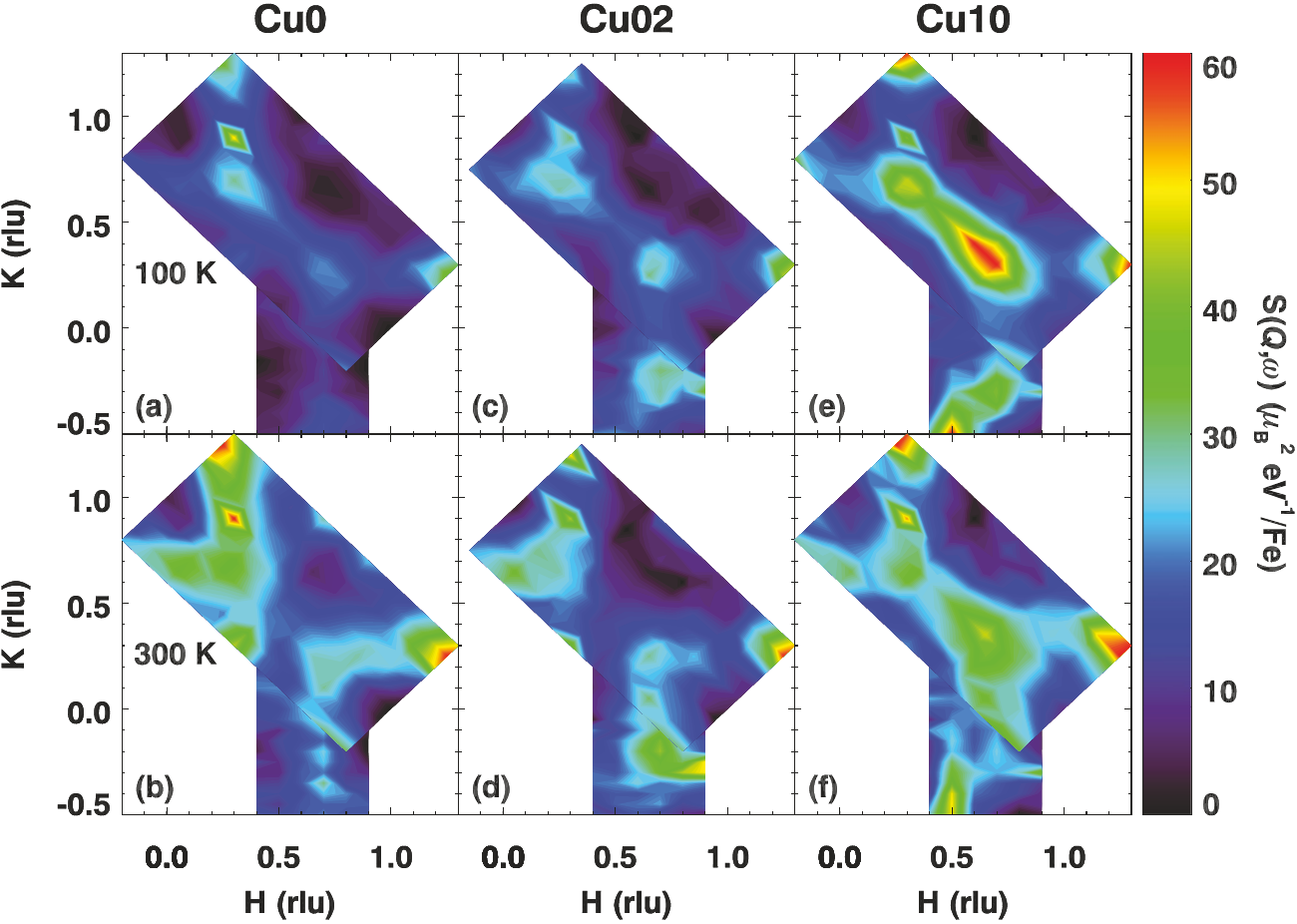}
\caption{Contour maps of Fe$_{0.98-z}$Cu$_z$Te$_{0.5}$Se$_{0.5}$ for a series of scans around (0.5 0.5) and (0.5 0) [(0.5 0) and (0.25 0.25) in unfolded notation, respectively] at a constant energy of 6\,meV at 100\,K (upper panel) and 300\,K (bottom panel) for different doping levels: $z = 0$, 0.02, and 0.1. Copyright by the American Physical Society. Reproduced from Ref.~\cite{Wen_Li_2013}.}
\label{fig:Wen}
\vspace{-0.2em}
\end{figure}

According to Anderson's theorem \cite{Anderson_1959}, NMI can break SC pairing only for an anisotropic gap. In cuprates, atoms of Zn were usually substituted for Cu, where the presence of only a few percents of Zn$^{2+}$ remarkably suppressed SC due to the presence of an anisotropic $d$-wave gap \cite{Balatsky_1994}. At the same time, early studies focused on Zn impurities in FBS reported contradictory results.

It was first reported that Zn doping in LaFeAsO$_{0.85}$ \cite{Guo_Shi_2010} and LaFeAsO$_{1-x}$F$_{x}$ \cite{Li_Tong_2010} severely suppressed the superconducting phase. A comparable result was obtained in the following work on K$_{0.8}$Fe$_{2-y-x}$Zn$_x$Se$_2$, where the substitution of nonmagnetic Zn for Fe substantially increased magnetic susceptibility of the material and strongly suppressed $T_\text{c}$ \cite{Tan_Zhang_2011}. In a further work \cite{Li_Guo_2015}, a more linear $T_\text{c}$ suppression was found for the Ba$_{0.5}$K$_{0.5}$Fe$_{2-2x}$Zn$_{2x}$As$_2$ superconductor, where the authors studied impurity effects of 3$d$-metal substitution by Mn, Co, Ni, Cu, Zn and Ru. Results of this work are shown in Fig.\,\ref{fig:Li}. One can see that SC is maintained robustly upon Ru doping and is increasingly suppressed by Co, Zn, Ni, Cu, and Mn impurities. The negligible suppression effect from Ru in Fig.\,\ref{fig:Li} is consistent with the `1111' system \cite{Lee_Satomi_2010, Sato_Kobayashi_2010}.

The local effect of a single interstitial impurity on a superconductor with the $s_\pm$ pairing symmetry \cite{Yu_Liu_2015} is shown in Fig.\,\ref{fig:NMI}. The spatial distribution of the self-consistent superconducting order parameter is given for the impurity potential. It is clear that a significant suppression of the SC pairing is locally present in the vicinity of the impurity~site.

An experimental attempt to highlight the role of impurity scattering for the pair-breaking mechanism as compared to the influence of carrier doping, structural distortions, or chemical pressure has been undertaken by K. Kirshenbaum \textit{et al.} \cite{KirshenbaumSaha12}. The authors compare the rate of $T_{\rm c}$ suppression induced by chemical substitution among numerous iron-based superconductors with the ThCr$_2$Si$_2$ (`122'-type) crystal structure with the experimental transport scattering rate. They reveal a universal dependence indicative of a common pair-breaking mechanism (Fig.~\ref{fig:Kirshenbaum}), letting them deduce an upper limit of 26\,K for the maximum $T_{\rm c}$ for all transition-metal-substituted 122 systems. On the other hand, they also note that the observed critical scattering rate is much weaker than expected for an $s_\pm$ multiband superconductor with strong interband scattering.

\begin{figure}[b!]
\includegraphics[width=\columnwidth]{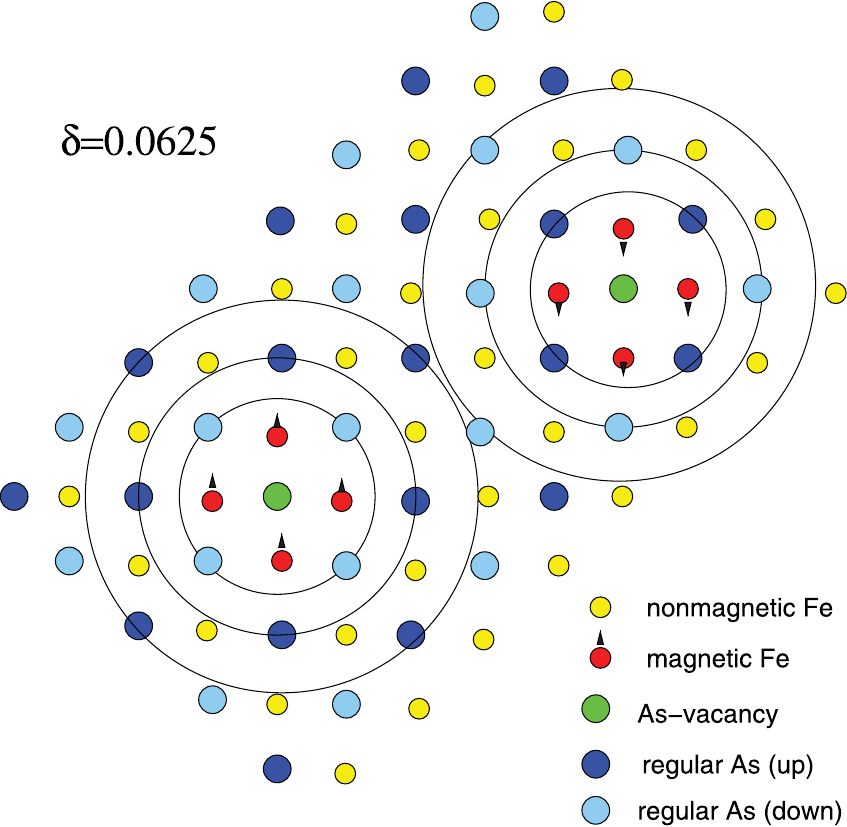}
\caption{Schematic structure of an FeAs block with two neighboring As-vacancies\cite{Grinenko_Kikoin_2011}. Copyright by the American Physical Society.}
\label{fig:Grinenko_Kikoin_2011}
\vspace{-0.2em}
\end{figure}

Recently NMI effects on spin excitations were studied in Fe$_{0.98-z}$Cu$_z$Te$_{0.5}$Se$_{0.5}$ by neutron-spectroscopy measurements \cite{Wen_Li_2013} upon transition metal substitution with weakly (non-) magnetic Cu. With increasing Cu substitution, the system is driven towards an insulator and the low-energy spin excitations of the system are enhanced (Fig.\,\ref{fig:Wen}, upper panels). Such results are unusual, as normally Cu doping suppresses spin waves, and can be explained by assuming that Cu inclusions introduce localization into the system and suppress the itinerancy.

In Fig.\,\ref{fig:Wen} we reproduce the momentum maps at a constant energy of 6\,meV at 100\,K (upper panel) and 300\,K (bottom panel) for different doping levels (from left to right: $z = 0$, 0.02, and 0.1). One can see that for 100\,K the spin excitations near (0.5 0.5) [(0.5 0) in the unfolded notation] become stronger with Cu doping, and there is no static magnetic order at (0.5 0) [(0.25 0.25) in the unfolded notation]. The scattering is incommensurate with the strongest scattering occurring at wave vectors displaced from (0.5 0.5). For each sample, as the temperature increases, the magnetic excitations become broader.

\begin{figure}[t!]
\begin{center}
\includegraphics[width=0.9\columnwidth]{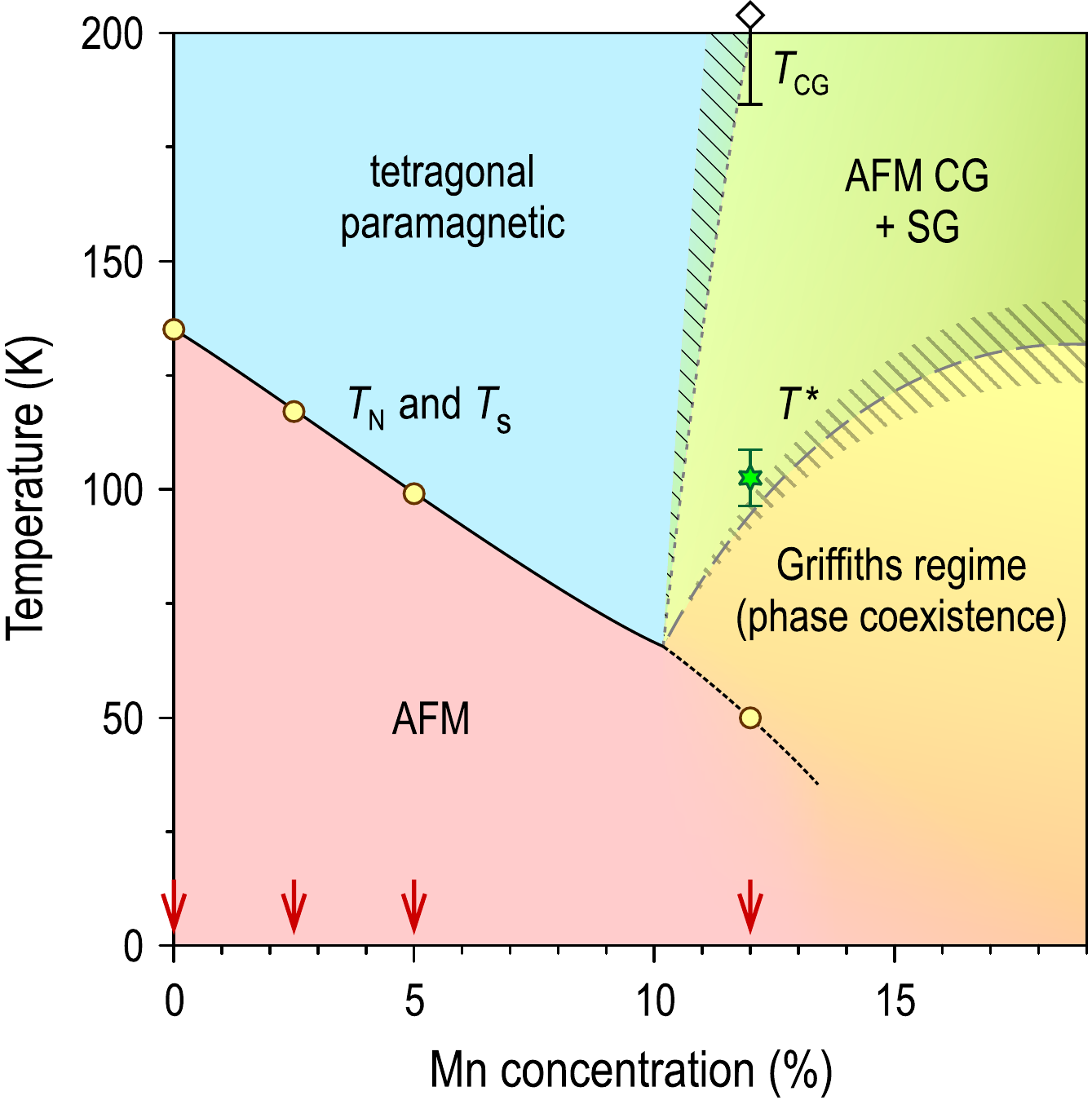}
\end{center}
\caption{Schematic phase diagram of BFMA, reprinted from Ref.~\cite{Inosov_Friemel_2013}, which summarizes the results of Refs.~\cite{Kim_2010, Kim_2011, Inosov_Friemel_2013}. The sample compositions investigated in \cite{Inosov_Friemel_2013} are indicated by arrows. Circles denote the SDW transition temperatures, $T_\text{N}$, below which the whole volume of the sample remains fully magnetic. $T_\text{N}$ was determined by transverse-field $\mu$SR spectroscopy in the same work. The diamond symbol marks the onset of the elastic neutron-scattering intensity, defined from neutron diffraction. Such intensity is associated with the formation of long-range magnetic correlations in the CG phase. The star symbol stands for $T_\text{CG}$, defined by the resistivity and $\mu$SR measurements. Copyright by the American Physical \mbox{Society}.\vspace{-5pt}}
\label{fig:Inosov_Friemel}
\end{figure}

Further, Kim \textit{et al.} compared the effects of transition-metal substitution by copper, which induces no superconductivity, to cobalt or nickel, which stabilize superconductivity, on the spin-fluctuation spectra of BaFe$_2$As$_2$ \cite{KimLamsal12}. They studied samples with 2.8\%~Cu and 4.7\%~Co concentrations, well below the optimal doping level, and observed that their spin fluctuation spectra are indistinguishable within the accuracy of the experiment. At the same time, in the elastic channel, Co- and Ni-substituted samples displayed incommensurate spin-density-wave (SDW) order, whereas the magnetic phase for Cu-substituted samples remained commensurate. Hence, the substitutional impurity effects (in contrast to the simple rigid-band doping concept) were found to be of major importance for controlling both the static magnetism and superconductivity.

\section{The effect of magnetic impurities}

The formation of a magnetic moment due to the presence of nonmagnetic arsenic vacancies in LaFeAsO was studied in Refs.~\cite{Kikoin_Drechsler_2015, Grinenko_Kikoin_2011} by the analysis of the static susceptibility and nuclear spin-lattice relaxation rate $1/T_1T$. An induced magnetic moment leads to the substantial enhancement of the paramagnetic susceptibility in both normal and superconducting states. Authors explained the origin of the moment formation by a strong enough $d\!-\!p$ hybridization between Fe $3d$ and As $4p$ orbitals. Arsenic defects remove the covalent bonds with $3d$ orbitals from four adjacent Fe ions, which leads to the formation of a vacancy-iron complex that is schematically depicted in Fig.\,\ref{fig:Grinenko_Kikoin_2011}. It leads to the charge transfer from Fe ions to As vacancies, which results in the local enhancement of the effective charge around Fe ions and formation of localized states with inherent uncompensated magnetic moments around defects.

The effect of both magnetic (Mn, Co, Ni) and nonmagnetic impurities (Ru, Cu, Zn) in BKFA was studied in Ref.~\cite{Li_Guo_2015}. One can see in Fig.\,\ref{fig:Li} that there is no direct correspondence between the SC pair breaking and the magnetic properties of the dopant atom. The localized magnetic Mn inclusions cause the strongest $T_\text{c}$ suppression among all studied impurities, which in the most extreme case of La-1111 became known as the ``poisoning effect'' \cite{HammerathBonfa14} that received a recent theoretical explanation in terms of cooperative impurity behavior driven by Coulomb interactions \cite{GastiasoroBernardini16}.

\begin{figure}[t!]
\mbox{\hspace{-0.01\columnwidth}\includegraphics[width=1.02\columnwidth]{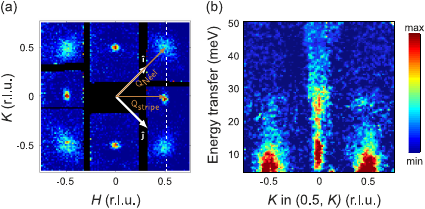}}
\caption{(a,b) The inelastic neutron scattering intensity at the conventional $(\pi,0)$ ($\textbf{Q}_\text{stripe}$) and the unconventional $(\pi,\pi)$ (\textbf{Q}$_\text{N\'{e}el}$) propagation vectors of 7.5\% doped Ba(Fe$_{1-x}$Mn$_x$)$_2$As$_2$. (b) The energy spectrum of $(\pi,0)$ and $(\pi,\pi)$ spin excitations plotted along the dashed line in panel (a). Reproduced from Ref.~\cite{Tucker_Pratt_2012}. Copyright by the American Physical Society.}
\label{fig:Tucker_Pratt}
\vspace{-0.2em}
\end{figure}

\begin{figure}[b!]
\includegraphics[width=\columnwidth]{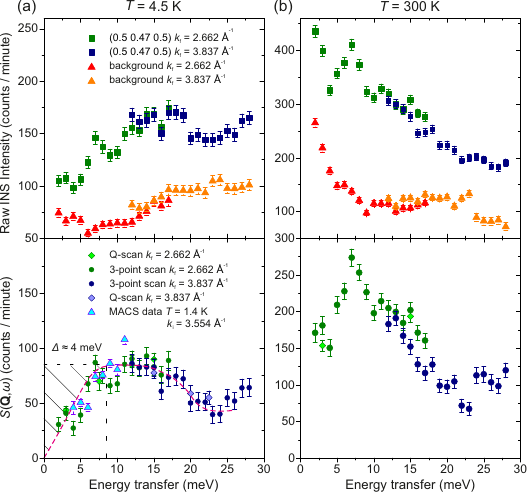}
\caption{Energy dependence of the INS intensity at $T=4$\,K and $T\!=\!300$\,K, which summarizes the results of our measurements at PUMA and MACS spectrometers. The top panels show raw unprocessed intensity. Bottom panels present background-subtracted intensity. Low-energy data at the bottom of panel (a) reveal the partial spin gap of $\sim$\,4\,meV (midpoint). Error bars represent one standard deviation.\vspace{-3pt}}
\label{fig:BFMA_our_1}
\end{figure}

\begin{figure}[t!]\vspace{-3pt}
\includegraphics[width=\columnwidth]{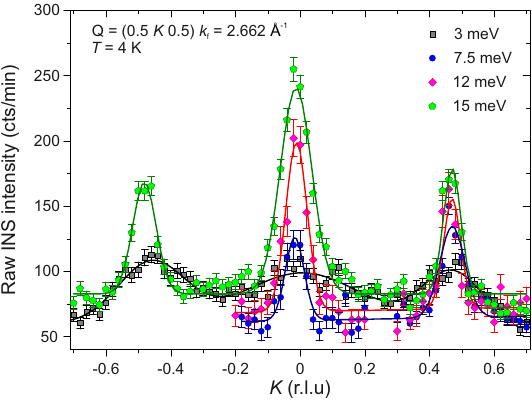}\vspace{1em}
\caption{Momentum scans along the (0.5 $K$ 0.5) direction at the zone boundary, crossing both the $(\pi,0)$ and $(\pi,\,\pi)$ excitation branches. The data were measured in the energy range 3\,--\,15\,meV at 4\,K. One can see the broadening and suppression of the signal at 3\,meV. The error bars stand for one standard deviation of the count rate.}
\label{fig:BFMA_our_2}\vspace{2em}
\mbox{\hspace{-0.02\columnwidth}\includegraphics[width=1.04\columnwidth]{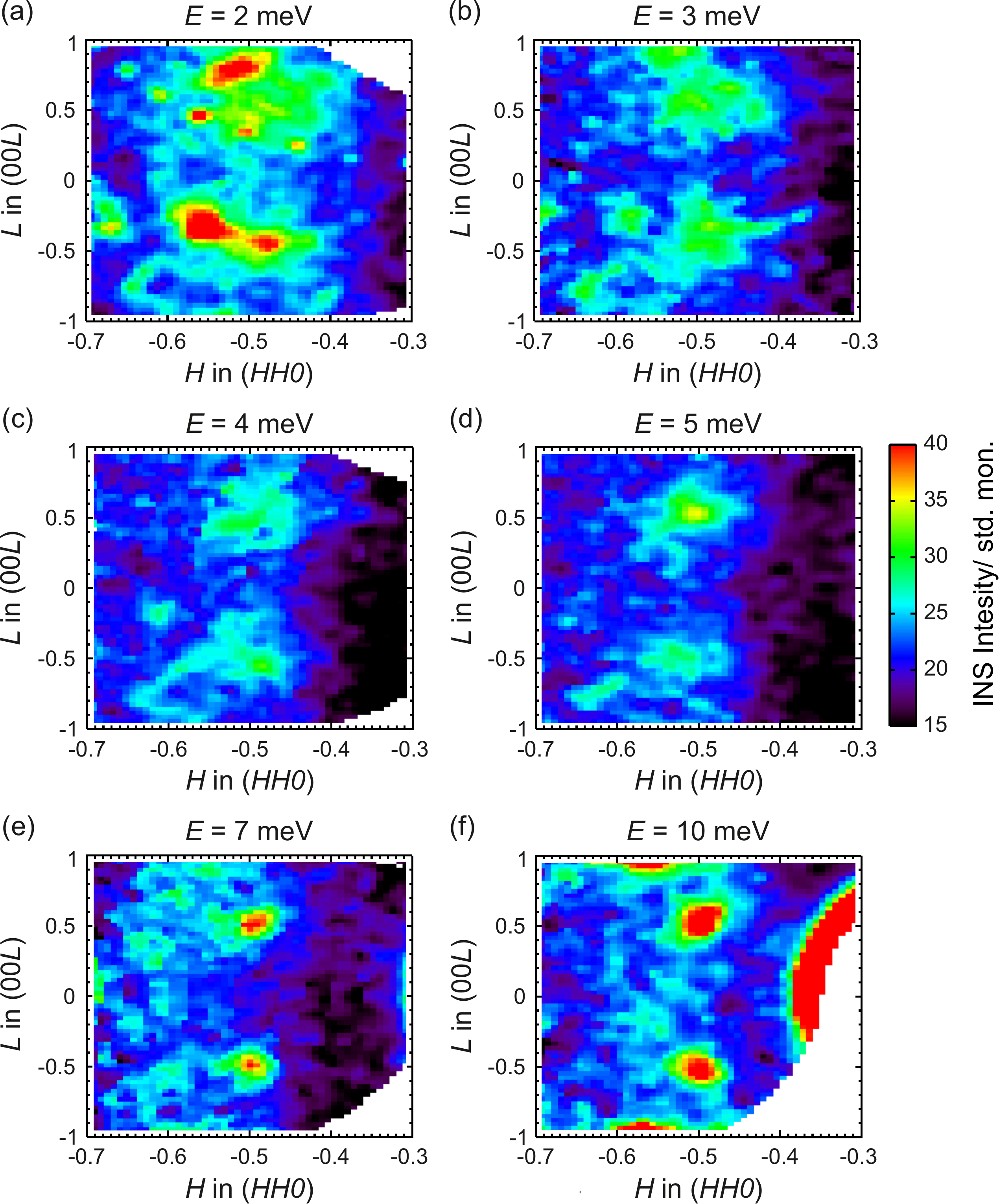}}\vspace{7pt}
\caption{Constant-energy maps of the magnetic intensity in the $(HHL)$ plane measured at the MACS spectrometer at $T$ = 1.6 K, showing two equivalent $(\pi,\pi)$ modes at $L=\pm\frac{1}{2}$ at several representative energies indicated above each panel. The sharp $L$-dependence of the magnetic intensity is indicative of highly three-dimensional AFM fluctuations.\vspace{-0.8em}}
\label{fig:BFMA_MACS}
\end{figure}

Indeed, magnetic Cr and Mn impurities substituted for Fe in `122'-type parent compounds represent unusual exceptions among other transition-metal dopants. First of all, they suppress the static AFM order without inducing SC \cite{Kim_2010, Kim_2011, Marty_2011, Inosov_Friemel_2013}, as seen in the phase diagram shown in Fig.~\ref{fig:Inosov_Friemel}. Recent NMR measurements \cite{Texier_Laplace_2012} indicated that this distinct behavior results from the localization of additional Mn holes, which prevents the change in the electron count within the conduction band, in contrast to Co or Ni dopants, but instead stabilizes local magnetic moments on the Mn sites. It was then found that these localized moments lead to the formation of unusual $(\pi,\pi)$ fluctuations in the excitation spectrum (Fig.~\ref{fig:Tucker_Pratt}), which were first explained by the formation of checkerboard-AFM nano\-regions stabilized by Mn impurities \cite{Tucker_Pratt_2012}.

\begin{figure}[t!]\vspace{-3pt}
\includegraphics[width=\columnwidth]{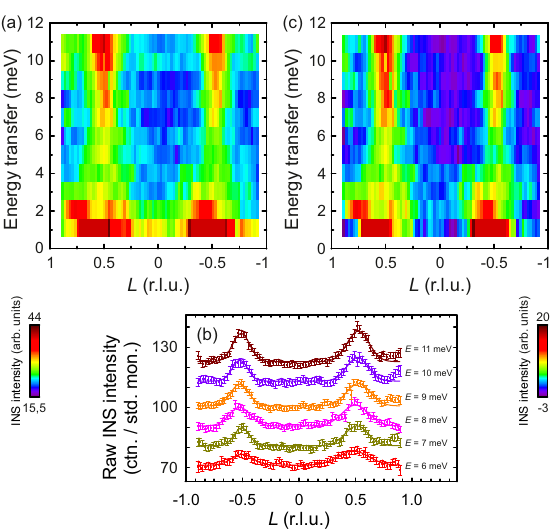}
\caption{Magnetic excitation spectra along the $(\frac{1}{2} \frac{1}{2} L)$ direction in reciprocal space, where unlike in the TOF data \cite{Tucker_Pratt_2012}, the energy transfer is not coupled to the out-of-plane momentum component $L$. (a) Color map of the total INS intensity from the MACS spectrometer, compiled from multiple momentum scans for different energies such as those shown in panel (b). (b) Out-of-plane dependence of the inelastic scattering signal for several representative energies, obtained by integrating the data in Fig.~\ref{fig:BFMA_MACS} around $H=\frac{1}{2}$ direction. The curves are spaced by 10 counts for clarity. Error bars represent one standard deviation of the count rate. (c) Color map of the background-subtracted INS intensity. Both panels (a) and (b) show the partial spin gap, which opens around 8\,meV, and an additional contamination from an acoustic phonon branch at low energies.}
\label{fig:BFMA_our_3}
\vspace{-0.2em}
\end{figure}

\begin{figure*}[t!]
\begin{center}\vspace{-1pt}
\includegraphics[width=0.75\textwidth]{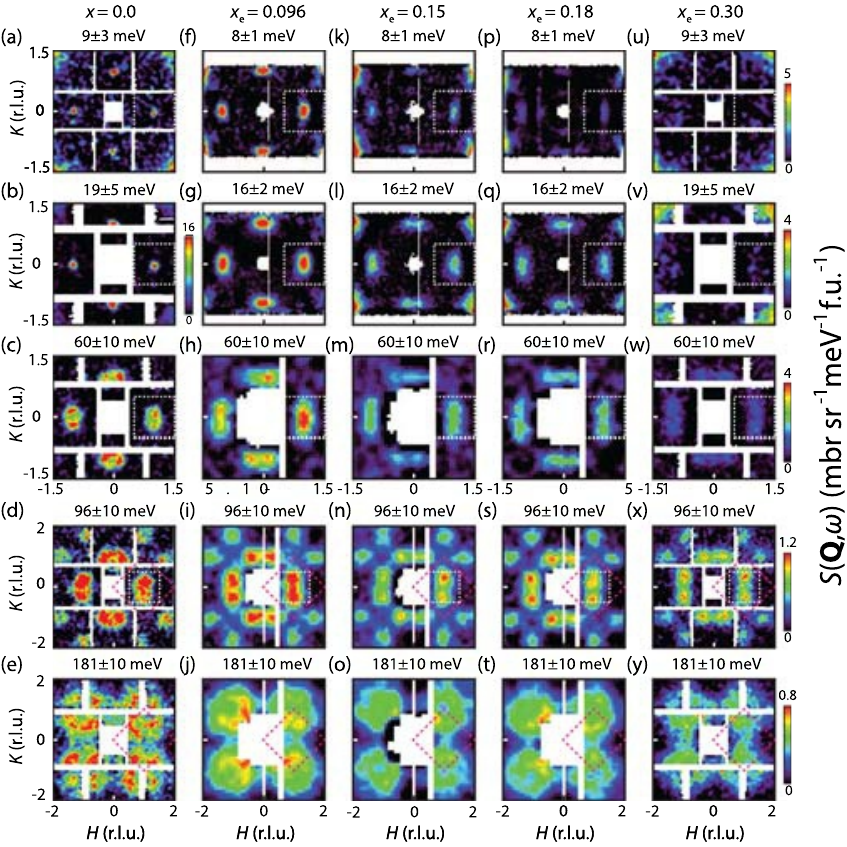}
\end{center}
\caption{Constant-energy maps of electron-doped BaFe$_{2-x}$Ni$_x$As$_2$ in the ($H K$) plane. Panels (a--e) illustrate spin excitations for the undoped compound BaFe$_2$As$_2$ \cite{Harriger_Luo_2011}, panels (f--j) for $x=0.096$, panels (k--o) for $x=0.15$, panels (p--t) for $x=0.18$ \cite{Luo_Yamani_2012}, and panels (u--y) for the nonsuperconducting $x=0.3$ sample \cite{Wang_2013}. Reproduced from Ref.~\cite{Dai_2015}. Copyright by the American Physical Society.\vspace{-3pt}}
\label{fig:Dai_2015}
\end{figure*}

The phase diagram for Ba(Fe$_{0.88}$Mn$_{0.12}$)$_2$As$_2$~(BFMA), reproduced from Ref.~\cite{Inosov_Friemel_2013} in Fig.~\ref{fig:Inosov_Friemel}, summarizes the results of $\mu$SR, resistivity, and neutron diffraction measurements \cite{Kim_2010, Kim_2011, Inosov_Friemel_2013}. It shows the temperature evolution of the AFM, cluster-glass (CG) \cite{Neto_Jones_2010, Dobrosavljevic_Miranda_2005, Westerkamp_Deppe_2009, Kassis_Vojta_2010}, and spin-glass (SG) phases \cite{Shell_Cowen_1982, Binder_Young_1986}. At intermediate temperatures, BFMA is characterized by the Griffiths regime of multiple coexisting phases, which is only present for Mn impurity concentration above 10\%. The formation of the CG phase is characterized by the presence of long-range AFM correlations between static magnetic clusters and defined by $T_{\text{CG}}{\rm}\approx210$\,K (the diamond symbol in the diagram) that is much higher than the ordering temperature of the undoped parent compound.

An INS study of Ba(Fe$_{1-x}$Mn$_x$)$_2$As$_2$ ($x = 7.5$\%) in Ref.~\cite{Tucker_Pratt_2012} has demonstrated the presence of the second branch of diffuse short-range spin fluctuations near the ($\pi$,\,$\pi$) wave vector, in contrast to other iron-based superconductors, where magnetic fluctuations are limited to $(\pi,0)$ and (0,\,$\pi$) nesting vectors. These fluctuations were interpreted as quasielastic, originating from checkerboard AFM fluctuations in the vicinity of Mn local moments. Both magnetic branches are illustrated in panels (a, b) in Fig.\,\ref{fig:Tucker_Pratt}. Panel (a) shows a constant-energy map of INS intensity obtained by integrating the time-of-flight (TOF) data within the energy window 5\,--\,15~meV. The data clearly show the presence of diffuse intensity around the ($\pi$,$\pi$) wave vector marked as $\mathbf{Q}_\text{N\'eel}$. Panel (b) shows the magnetic spectrum along the dashed line in panel (a). At low energies, the TOF data are limited by 5\,meV, and hence leave an open question about the presence of a low-energy spin gap. Further, due to the coupling between the energy and $L$ component of momentum in this experimental geometry, the data in panel (b) do not reflect the actual energy dependence of the scattering function. On the other hand, triple-axis data from Ref.~\cite{Tucker_Pratt_2012} demonstrate finite magnetic intensity at 3\,meV with a strong $L$ dependence, which can only be reconciled with a very small or partially developed spin gap.

To get a clearer picture of the peculiar ($\pi$,$\pi$) excitations that are uncommon among other iron pnictides, we performed follow-up INS measurements on a similar sample with 12\% Mn doping, the same that we investigated previously in Ref.~\cite{Inosov_Friemel_2013}. Our INS experiments were carried out at the thermal-neutron triple-axis spectrometer (TAS) PUMA at FRM-II (MLZ, Garching) \cite{Sobolev_Park_2015} and the cold-neutron multi-axis crystal spectrometer (MACS) at NIST (Gaithersburg, MD) \cite{Rodriguez_Adler_2008}. We used an assembly of coaligned Ba(Fe$_{0.88}$Mn$_{0.12}$)$_2$As$_2$ single crystals with a total mass of $\sim$\,1~g, which were grown from self-flux in zirconia crucibles sealed in quartz ampoules under argon atmosphere, as described elsewhere \cite{Liu_Sun_2010}. The lattice parameters of our sample, as measured at MACS during the sample alignment at the base temperature of 1.6\,K, were $a/\sqrt{2} = b/\sqrt{2} \approx 2.73$\,\AA\ and $c/2 \approx 6.33$\,\AA.

For measurements at the MACS spectrometer, the neutron final energy was fixed at $E_\text{f}=5$\,meV or $k_\text{f}=1.55$\,\AA$^{-1}\!$.\, The sample was mounted into a standard cryostat with (1\,1\,0) and (0\,0\,1) directions in the scattering plane. For PUMA measurements, the sample was aligned in the $(H K H)$ scattering plane using the (1\,0\,1) and (0\,1\,0) Bragg reflections, and the incident neutron wave vector was fixed at $k_\text{f}=2.662$ or $3.84$~\AA$^{-1}$ (14.68~meV and 30.55~meV, respectively). To avoid higher-order neutron contamination, we used a cold beryllium filter for MACS and two pyrolytic graphite filters for PUMA. The sample environment consisted of a closed-cycle cryostat at PUMA and an orange-type cryostat at MACS.

\begin{figure}[b!]
\includegraphics[width=\columnwidth]{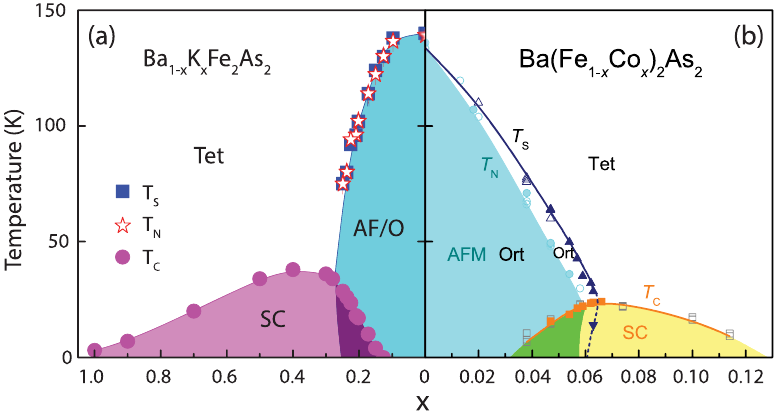}
\caption{The structural and the magnetic phase diagrams of electron- and hole-doped BaFe$_2$As$_2$. Panel (a) illustrates the structural and magnetic phase transitions for hole-doped Ba$_{1-x}$K$_x$Fe$_2$As$_2$, obtained from neutron powder diffraction experiments \cite{Avci_Chmaissem_2012}. Panel (b) summarizes electron doping of Ba(Fe$_{1-x}$Co$_x$)$_2$As$_2$, determined from x-ray and neutron diffraction experiments \cite{Nandi_2010}. Copyright by the American Physical Society.}
\label{fig:Avci_Nandi}
\end{figure}

We first present the energy dependence of INS intensity measured at PUMA. In Fig.\,\ref{fig:BFMA_our_1} (top panels) we show the raw data and background at $T=4.5$\,K and $T=300$\,K for two different values of $k_\text{f}$. The energy dependence of the scattering function, $S(\textbf{Q},\,\omega)$, which we obtained by background substraction, is shown in the lower panels for both low (4\,K) and high (300\,K) temperatures. Data from the MACS spectrometer are added to the plot as blue triangles at the bottom of panel (a). These data conform with results from PUMA spectrometer and follow the overall energy dependence, which is shown by the dashed line. Results from both spectrometers reveal the partial energy gap with a midpoint around 4\,meV and an onset below 7\,meV (hatched region). Despite the presence of this partial gap, finite magnetic intensity persists down to the lowest measurable energies, which is qualitatively consistent with the observation of an $L$-dependent signal at 3\,meV in Ref.\,\cite{Tucker_Pratt_2012}. The magnitude of this gap is approximately twice smaller as compared to the anisotropy gap measured on the same sample for the conventional $(\pi,0)$ branch of magnetic excitations \cite{Inosov_Friemel_2013}. The presence of the gap clearly demonstrates that the diffuse ($\pi$,\,$\pi$) excitations at 12\% Mn concentration are no longer quasielastic, but represent another spin-wave-like collective mode of short-range dynamic fluctuations. It could originate from the static magnetic clusters surrounding the Mn ions \cite{Inosov_Friemel_2013}, which start forming long-range AFM correlations below $T_{\rm CG}$ while filling only a fraction of the sample volume, whereas the rest of the sample exhibits the conventional $(\pi,0)$ magnetic correlations.

\begin{figure}[t!]
\hspace{-10pt}\includegraphics[width=1.05\columnwidth]{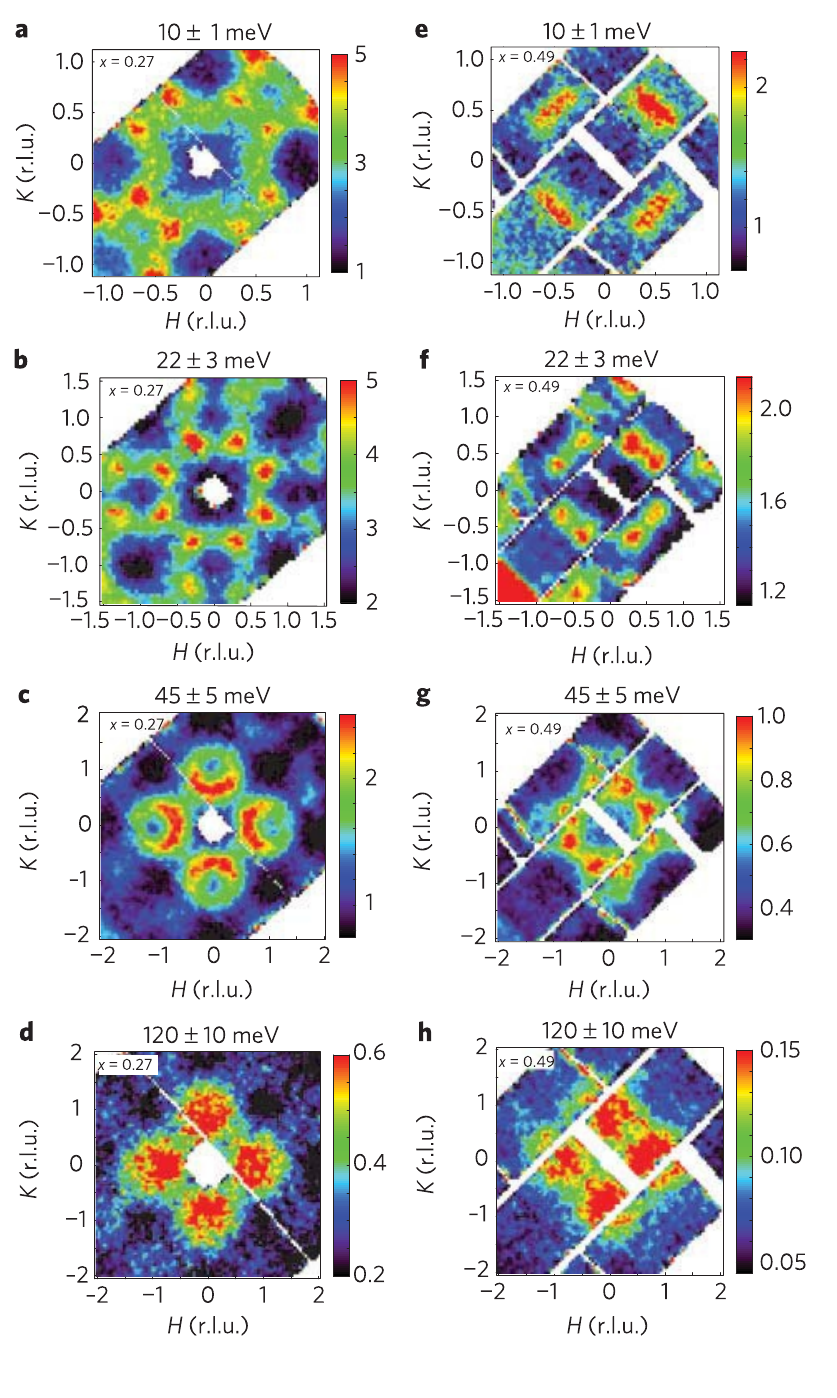}
\caption{Constant-energy maps of spin excitations in the ($H K$) plane of nonsuperconducting Fe$_{1+y}$Te$_{0.73}$Se$_{0.27}$ and superconducting Fe$_{1+y}$Te$_{0.51}$Se$_{0.49}$ at different representative energies. Reproduced from Ref.~\cite{Lumsden_2010}. Copyright by the Macmillan Publishers Ltd.}
\label{fig:Lumsden_Christianson_2010}
\vspace{-0.2em}
\end{figure}

Low-temperature scans along the (0.5 $K$ 0.5) direction, crossing both branches of magnetic excitations at several representative energies between 3 and 15~meV, are shown in Fig.~\ref{fig:BFMA_our_2}. One can see the broadening of the signal at 3~meV within the spin-gap region, yet the interpretation of this effect is not straightforward because of the possible contamination of the magnetic signal by acoustic phonons in the vicinity of the structural Bragg peak. The centers of all peaks are slightly shifted from their expected positions, which could be explained by resolution effects or a slight misalignment of the crystal mosaic during the measurement.

As the next step, we reveal the true out-of-plane momentum and energy dependence of the excitation spectrum along the (0.5 0.5 $L$) direction. In the data obtained with the MACS spectrometer, unlike in the TOF data, the energy transfer was not coupled to the out-of-plane component of the momentum $L$.

\begin{figure}[t!]
\includegraphics[width=\columnwidth]{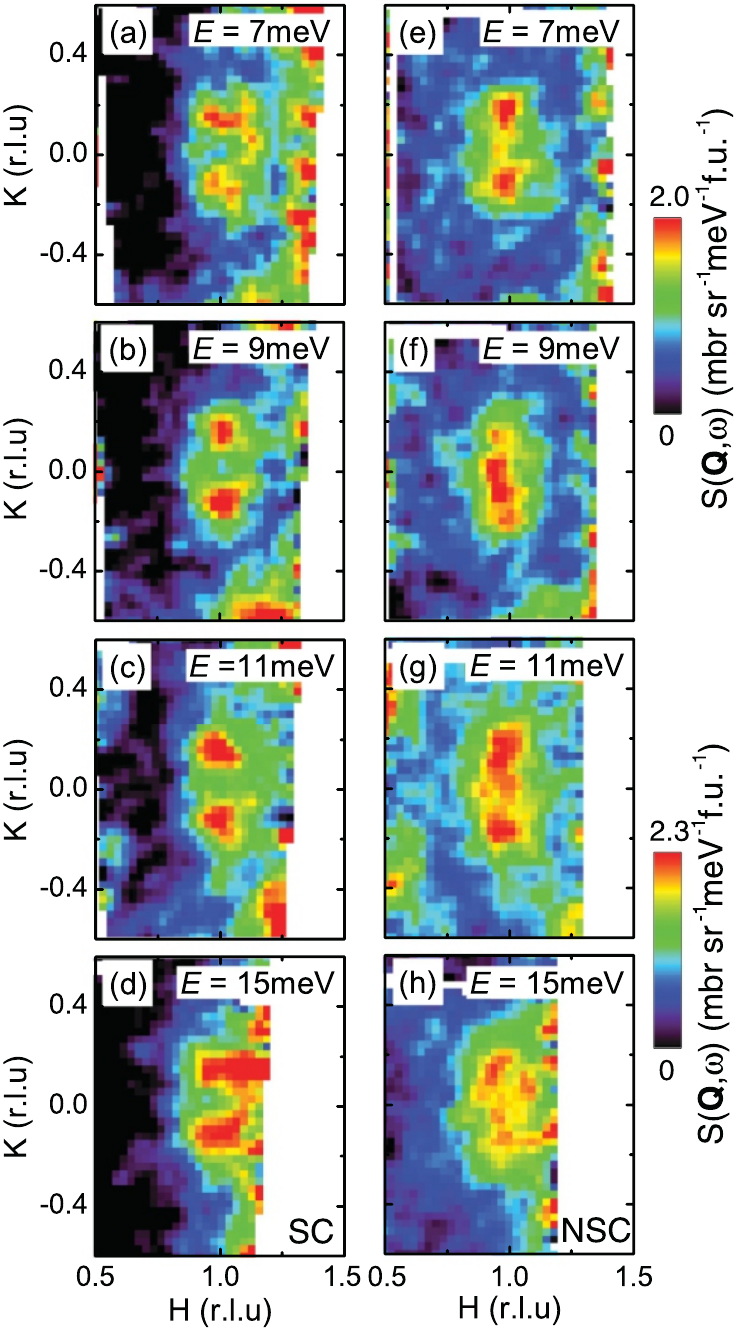}
\caption{Constant-energy maps in the ($H K$) plane of the superconducting LiFeAs [panels (a)-(d)] and nonsuperconducting Li$_{0.94}$FeAs [panels (e)-(h)] made for several representative energies. Reproduced from Ref.~\cite{Wang_Wang_2012}. Copyright by the American Physical Society.}
\label{fig:Wang_Wang_2012}
\vspace{-0.2em}
\end{figure}

\begin{figure}[t!]
\includegraphics[width=\columnwidth]{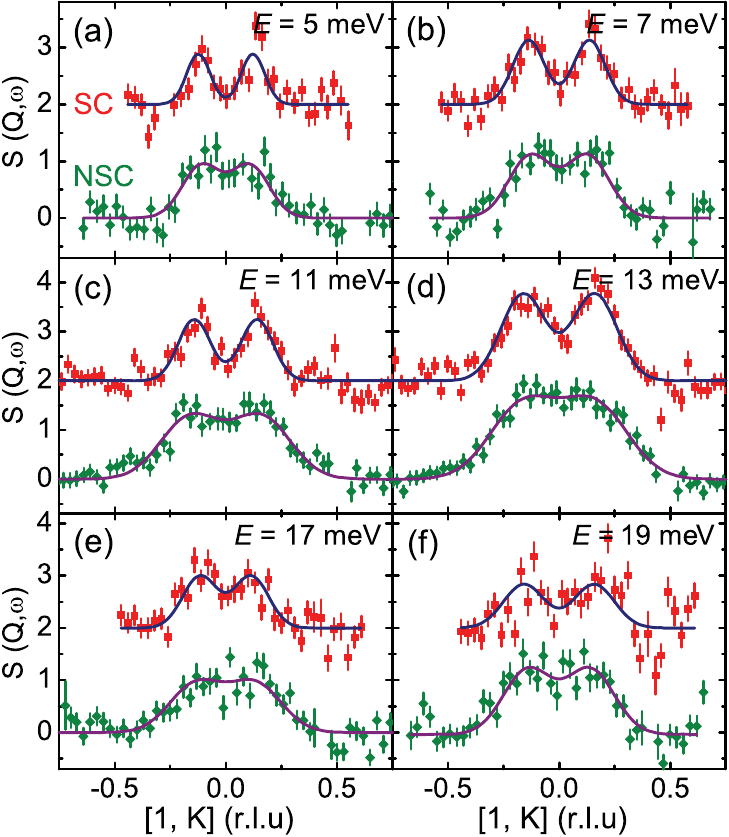}
\caption{Constant-energy cuts showing spin excitations along the (1 $K$) direction. The red and green data points correspond to the LiFeAs superconductor and the nonsuperconducting Li$_{0.94}$FeAs, respectively. Solid lines denote Gaussian fits. Reproduced from Ref.~\cite{Wang_Wang_2012}. Copyright by the American Physical Society.}
\label{fig:Wang_Wang_2012_2}
\vspace{-1.2em}
\end{figure}

These measurements were done by collecting a series of constant-energy maps in the $(HHL)$ scattering plane as shown in Fig.~\ref{fig:BFMA_MACS}. They demonstrate the presence of magnetic modes at the equivalent $(0.5~0.5~\overline{0.5})$ and $(0.5~0.5~0.5)$ wave vectors with a sharp $L$-dependence, which clearly indicates the three-dimensional character of these $(\pi,\pi)$ spin fluctuations. Furthermore, one can see a minimum of intensity around 4\,meV within the partial spin gap. At even lower energies ($\leq 3$ meV), the magnetic signal is contaminated presumably by an acoustic phonon mode. The same data were used to extract the energy-momentum dependence of the signal along the out-of-plane momentum direction by integrating out the in-plane momentum in the vicinity of the ordering vector, as shown in Fig.\,\ref{fig:BFMA_our_3} (a--c). Panel (a) shows a color map of the raw INS intensity compiled from multiple momentum scans for different energies. Several such scans are presented in panel (b), which shows unprocessed scattered intensity along the (0.5 0.5 $L$) direction, spaced by 10 counts on the vertical axis for clarity. Panel (c) shows the same data as in panel (a) after linear background subtraction. Both color maps reveal the partial spin gap with an onset around 8 meV, seen despite the phonon contamination at low energies. We observe that the low-energy magnetic intensity is peaked only at half-integer $L$ values and fully vanishes in the $L=0$ plane, which indicates strong $c$-axis correlations of the fluctuating magnetic moments, in agreement with the earlier conclusions of Tucker \textit{et al.} for the 7.5\% Mn concentration ~\cite{Tucker_Pratt_2012}.\vspace{-3pt}

\begin{figure}[t!]
\includegraphics[width=\columnwidth]{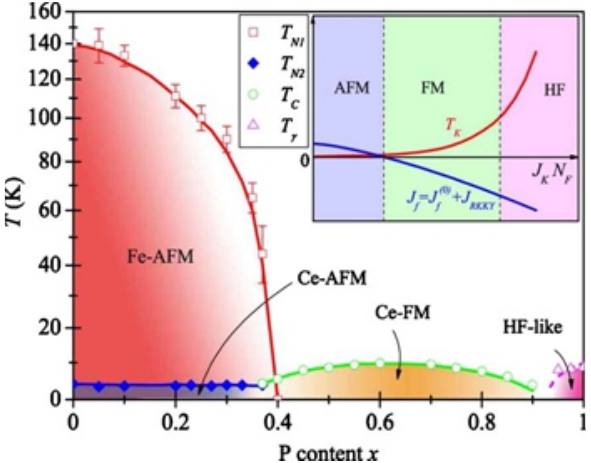}
\caption{Electronic phase diagram of CeFeAs$_{1-x}$P$_x$O with $0\leq x\leq1$. The shaded area in red (Fe moments) and blue (Ce moments) at the left part of the phase diagram denote AFM ordering. The pink area at the right corner is a nonmagnetic state with heavy-fermion behavior. The middle part of the diagram illustrates ferromagnetic ordering (Ce-FM). Reproduced from Ref.~\cite{Luo_2010}. Copyright by the American Physical Society.}
\label{fig:Luo_Li_2010}
\vspace{-0.5em}
\end{figure}

\section{Impurity effects of dopants}
Charge dopants introduced as substitutional impurities in iron pnictides usually lead to the suppression of static magnetic order and emergence of the superconducting phase.
Typical FBS families, such as Ba(Fe$_{1-x}M_x$)$_2$As$_2$ ($M$\,=\,Ni,~Co, etc.), exhibit a tetragonal-to-orthorhombic transition, which is usually separated from the magnetic phase transition at $T_\text{N}$ and occurs at a slightly higher temperature, $T_\text{s}$, as a result of an electron-nematic instability. By doping the sample, one can effectively change these representative temperatures. At the same time, it remains a question of debate whether the impurity-induced electron scattering or doping-induced changes in the electronic structure (and, consequently, Fermi-surface nesting properties) is the main underlying reason for the experimentally observed phase diagram.\smallskip

\begin{figure}[b!]
\includegraphics[width=\columnwidth]{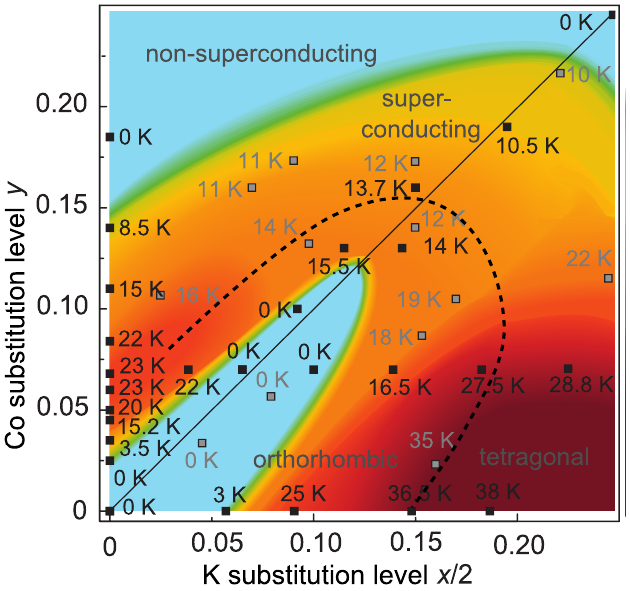}
\caption{Phase diagram of Ba$_{1-x}$K$_x$(Fe$_{1-y}$Co$_y$)$_2$As vs. two dopant concentrations. Blue color represents the nonsuperconducting ground state and superconductivity is
shown from yellow ($T_\text{c}=10$\,K) to dark red ($T_\text{c}=40$\,K). Dashed line denotes the border for the existence of the orthorhombic
phase. Reproduced from Ref.~\cite{Goltz_Zimth_2014}. Copyright by the American Physical Society.}
\label{fig:Goltz_Zimth_2014_1}
\end{figure}

\subsection{Hole and electron doping}

The evolution of structural and magnetic phase transitions for the typical `122'-type compound, such as BaFe$_2$As$_2$, upon electron and hole doping is depicted in Fig.\,\ref{fig:Avci_Nandi}. The left panel (a) shows hole doping of Ba$_{1-x}$K$_x$Fe$_2$As$_2$ from neutron powder diffraction experiments \cite{Avci_Chmaissem_2012}. The right panel (b) illustrates x-ray and neutron diffraction measurements on the electron-doped Ba(Fe$_{1-x}$Co$_x$)$_2$As$_2$. At zero doping, their common BaFe$_2$As$_2$ parent compound exhibits a magnetic and tetragonal-to-orthorhombic structural transitions, which occur almost at the same temperature of $\sim$\,138\,K. Upon electron doping with Co, shown in Fig.\,\ref{fig:Avci_Nandi}\,(b), structural and magnetic transitions become gradually separated. The coexistence of superconductivity and magnetism is shown as the shaded area in the underdoped regime at both phase diagrams. In contrast to hole doping, for optimal electron doping the system enters the reentrant tetragonal phase.

In the presence of charge doping, the spectrum of spin dynamics changes. In moderately electron-doped `122' systems, this change represents a reduction and smearing of the anisotropy gap \cite{Tucker_2014} concomitant with the reduction of dimensionality of anisotropic spin excitations \cite{Harriger_2009}. The signal broadens and becomes more diffusive \cite{Tucker_2014}. With higher electron doping, the local gap anisotropy develops into a spin-glass phase with a smeared N\'{e}el transition \cite{Lu_Tam_2014}.

\begin{figure}[t!]
\includegraphics[width=\columnwidth]{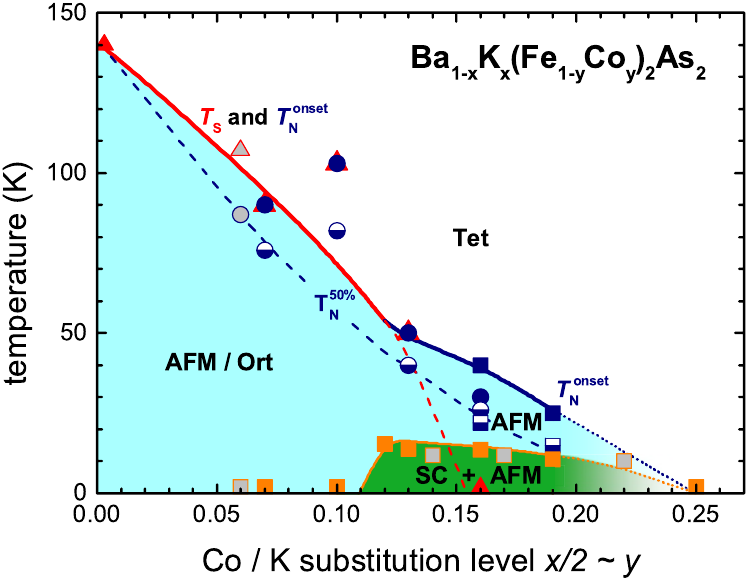}
\caption{Phase diagram of charge-compensated Ba$_{1-x}$K$_x$(Fe$_{1-y}$Co$_y$)$_2$As$_2$ along the $x/2 \approx y$ diagonal, illustrating the coexistence of the AFM state with superconductivity. Reproduced from Ref.~\cite{Goltz_Zimth_2014}. Copyright by the American Physical Society.}
\label{fig:Goltz_Zimth_2014_2}
\end{figure}

One can observe the evolution of spin excitations in the ($H$\,$K$) plane at different energies as a function of the electron doping for BaFe$_{2-x}$Ni$_x$As$_2$ that is shown in Fig.\,\ref{fig:Dai_2015}. In the undoped compound BaFe$_2$As$_2$, due to the presence of a spin gap below $\sim$\,15\,meV, the signal at the conventional ordering vector ($\pm$0.5 0) and its twin (0 $\pm$0.5) around $E=9$\,meV has low intensity [denoted as ($\pm$1 0) and (0 $\pm$1) in Fig.\,\ref{fig:Dai_2015}\,(a)]. For optimally electron-doped $x=0.096$, the spin gap is suppressed, which leads to the enhancement of low-energy spin excitations [Fig.\,\ref{fig:Dai_2015}\,(f)]. For the electron-overdoped range $x=0.15$ and $x=0.18$, low-energy spin fluctuations become weaker and more elongated [Fig.\,\ref{fig:Dai_2015}\,(k,p)]. For a nonsuperconducting sample with $x=0.3$, the spin excitation spectrum up to $\sim$\,20\,meV [Fig.\,\ref{fig:Dai_2015}\,(u--v)] is completely suppressed, which can be explained by the presence of a large spin gap. At the same time, high-energy spin excitations remain mostly unchanged upon electron doping \cite{Harriger_Luo_2011, Luo_Yamani_2012, Wang_2013, Dai_2015}.

The evolution of spin excitations at different energies for the nonsuperconducting sample Fe$_{1+y}$Te$_{0.73}$Se$_{0.27}$ and superconducting Fe$_{1+y}$Te$_{0.51}$Se$_{0.49}$ is represented in Fig.\,\ref{fig:Lumsden_Christianson_2010} (left and right columns, respectively) \cite{Lumsden_2010}. The low-energy excitations of the nonsuperconducting sample [panels (a) and (b) in Fig.\,\ref{fig:Lumsden_Christianson_2010}] are centered at the transversely incommensurate positions near ($\pm$0.5~$\pm$0.5) [or ($\pm$0.5~0)$/$(0~$\pm$0.5) in the unfolded notation]. At higher energies, spin excitations become fourfold symmetric and centered around ($\pm$1~0) and (0~$\pm$1) [or ($\pm$0.5~$\pm$0.5) in the unfolded notation, see panels (c) and (d)]. On the other hand, for the superconducting compound the low-energy excitations are characterized by elongated spin excitations around ($\pm$0.5~$\pm$0.5) [panels (e) and (f)]. The high-energy spectrum [panels (g) and (h)] shows no significant differences as compared to the nonsuperconducting sample [panels (c) and (d)]. One can therefore conclude that charge doping affects only the low-energy part of the spectrum.\smallskip\smallskip

\subsection{Vacancies, defects and isovalent substitution}

Isovalent substitution, such as replacing As with P \cite{Jiang_2009} or Fe with Ru \cite{Kim_2011} in BaFe$_2$As$_2$, offers other interesting examples of impurity effects. The peculiarity of the phase diagram of BaFe$_2$(As$_{1-x}$P$_x$)$_2$ is the presence of a magnetic quantum critical point, which was proven by transport and London penetration depth measurements \cite{Shibauchi_2014}. At the same time, more recent NMR, x-ray and neutron diffraction measurements in Ref.~\cite{Hu_2015} report that structural and magnetic transitions vanish like in the electron-doped iron pnictides with an avoided quantum critical point.

\begin{figure}[b!]
\includegraphics[width=\columnwidth]{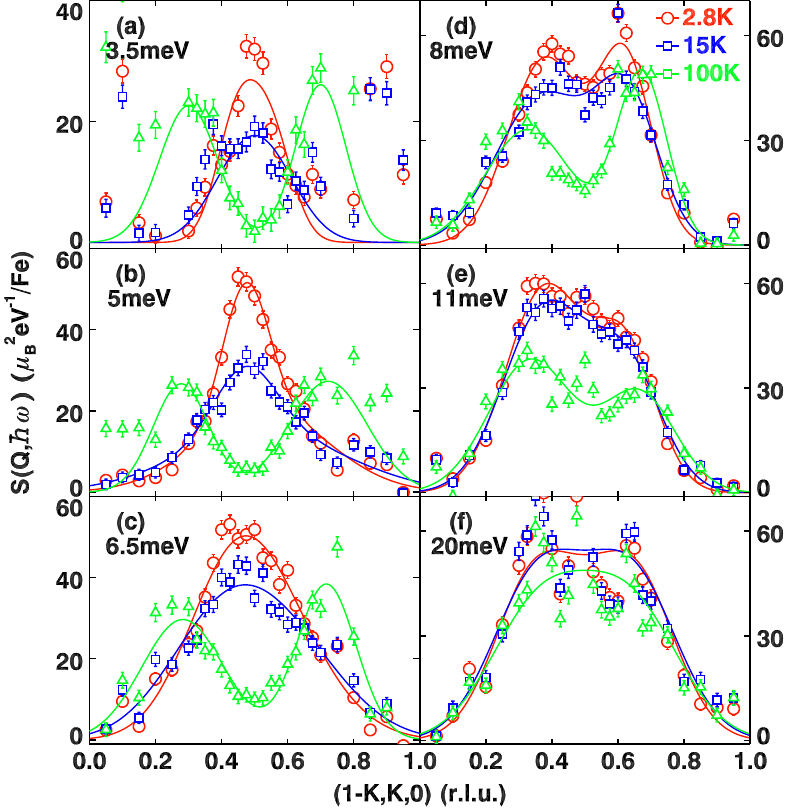}
\caption{Wave vector dependence of the magnetic scattering intensity in Fe$_{0.96}$Ni$_{0.04}$Te$_{0.5}$Se$_{0.5}$ along the transverse direction near (0.5 0.5) [equivalent to (0.5 0) in the unfolded notation] at different temperatures: 2.8\,K (red cirles), 15\,K (blue squares) and 100\,K (green triangles). Panels (a)-(f) illustrate different representative energies. Solid lines are guides to the eyes. Reproduced from Ref.~\cite{Xu_Wen_2012}. Copyright by the American Physical Society.}
\label{fig:Xu_Wen}
\vspace{-0.2em}
\end{figure}

The role of vacancies can be reviewed by the example of LiFeAs, a superconductor without any static magnetic order \cite{Tapp_2008}, where Li deficiency in Li$_{1-x}$FeAs suppresses superconductivity \cite{Wang_Wang_2012, Qureshi_2012, Qureshi_2014}. Early ARPES measurements reveal poor Fermi surface nesting between the $\Gamma$ and $M$ points \cite{Borisenko_2010}. This is considered to be the main reason for the absence of magnetic ordering. Further, INS measurements performed on both superconducting LiFeAs compound and nonsuperconducting sample with Li vacancies revealed transversely incommensurate spin excitations, arising from the quasiparticle excitations between mismatched hole and electron Fermi surfaces \cite{Wang_Wang_2012, Qureshi_2012, Qureshi_2014}. In Fig.\,\ref{fig:Wang_Wang_2012} we show constant-energy maps of spin excitations in the ($H K$) plane reproduced from Ref. \cite{Wang_Wang_2012}. Panels (a)--(d) illustrate data from LiFeAs, panels (e)--(h) show results from Li$_{0.94}$FeAs. Both samples reveal the incommensurate spin excitations at all energies. However, the incommensurate spin excitations of LiFeAs have better defined peaks than magnetic excitations in the nonsuperconducting sample. One can see this difference in Fig.\,\ref{fig:Wang_Wang_2012_2}, where constant-energy cuts through the momentum-dependent magnetic scattering function are shown. It is worth to note that spin excitations in both Fig.\,\ref{fig:Wang_Wang_2012} and Fig.\,\ref{fig:Wang_Wang_2012_2} are weakly energy dependent, only showing a broadening effect towards higher energies.

Another interesting example of isovalent doping is CeFeAs$_{1-x}$P$_x$O, where arsenic is substituted by phosphorus. In this system, static AFM order can be suppressed without superconductivity \cite{Cruz_2010}. Authors performed detailed neutron diffraction studies on the powder samples and revealed the presence of a magnetic quantum critical point. Further transport and susceptibility measurements \cite{Luo_2010} revealed that superconductivity does not appear in the entire phase diagram. Such peculiarity can possibly arise due to heavy-fermion properties of Ce\,4f electrons. The resulting phase diagram is reproduced in Fig.\,\ref{fig:Luo_Li_2010}.\smallskip

\begin{figure}[t!]
\includegraphics[width=\columnwidth]{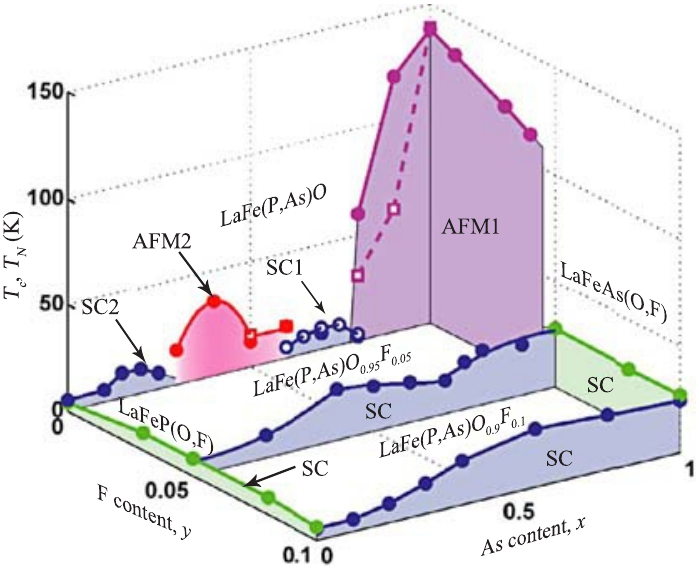}
\caption{The phase diagram of LaFe(As$_{1-x}$P$_x$)O$_{1-y}$F$_y$. Reproduced from Ref.~\cite{Lai_2014}. Copyright by the American Physical Society.}
\label{fig:Lai_Takemori}
\vspace{-0.2em}
\end{figure}

\subsection{Codoping}

Simple chemical substitution leads to a number of different effects such as charge doping, impurity scattering, or chemical pressure, which are difficult to separate. To decouple the charge doping from other impurity effects, some iron pnictides were codoped with different concentrations of two elements to preserve the constant charge carrier density while varying the impurity concentration. One of the most studied compounds in this field is Ba-122 codoped with Co and K \cite{Goltz_Zimth_2014}. This work demonstrated the presence of superconductivity up to $T_\text{c}=15.5$\,K, which can be evoked even in the charge-compensated regime in coexistence with static AFM order. These results are shown in Fig.\,\ref{fig:Goltz_Zimth_2014_1} and Fig.\,\ref{fig:Goltz_Zimth_2014_2}, which summarize the dependence of $T_\text{c}$ on dopant concentrations and the suppression of magnetic order along the line of charge compensation.

Another example of codoping is the introduction of magnetic Ni into iron chalcogenides with the statistical mixture of Se and Te, resulting in local disorder, which was studied in Ref.~\cite{Xu_Wen_2012}. Based on INS measurements, the authors observed an incommensurability of low-energy spin excitations in Fe$_{0.96}$Ni$_{0.04}$Te$_{0.5}$Se$_{0.5}$. These results are illustrated in Fig.\,\ref{fig:Xu_Wen}, where we show the momentum dependence of the magnetic scattering function $S(\textbf{Q},\omega)$ for several representative energies. INS data obtained at different temperatures are depicted in red, blue, and green datasets, respectively. As one can clearly see, at low energies [panels (a)-(c)] a single commensurate peak below $T_\text{c}$ transforms into a pair of well-resolved incommensurate peaks around 100\,K. For higher energies in panels (d)-(f), the splitting of two peaks is only partially reduced with temperature, but the overall trend persists.

The codoping of La-1111 with fluorine and phosphorus \cite{Lai_2014} reveals a peculiar phase diagram of LaFe(As$_{1-x}$P$_x$)O$_{1-y}$F$_y$, reproduced in Fig.\,\ref{fig:Lai_Takemori}. A new AFM phase was reported for zero fluorine concentration and the phosphorus doping range $0.4\leq x\leq0.7$, where two superconducting domes are separated by an AFM region with a maximum at $T_\text{N}=35$\,K \cite{Mukuda_2014}. Upon fluorine doping, this AFM phase is suppressed, and the two superconducting domes merge into a single one with two maxima of $T_{\rm c}$ as a function of the As content. Such a trend suggests the presence of an unusual AFM quantum critical point at low fluorine concentration $0<y_0<0.05$ and a certain phosphorus concentration $x_0$, where any change of $x$ with respect to $x_0$ will lead to the increase of $T_\text{c}$.\smallskip\smallskip

{\subsection*{\small Acknowledgements}\small~\,The authors acknowledge financial support by the German Research Foundation (DFG) within the priority program SPP\,1458/2 (Grant No.~IN209/1-2) and the Graduiertenkolleg GRK\,1621 at the TU Dresden. This work utilized facilities supported in part by the US National Science Foundation under Agreement No. DMR-1508249.
}

\end{document}